\documentclass[apj]{emulateapj}

\bibpunct{(}{)}{;}{a}{}{,}

\slugcomment{Accepted to the ApJ}

\shorttitle{A Lens Model for a SMG At $z=4.243$}
\shortauthors{Bussmann et al.}

\newcommand{\tnm}{\tablenotemark}
\newcommand{\tnt}{\tablenotetext}

\begin{document}

\title{A Detailed Gravitational Lens Model Based on Submillimeter
Array\footnote{Some of the data presented herein were obtained at the
Submillimeter Array, which is a joint project between the Smithsonian
Astrophysical Observatory and the Academia Sinica Institute of Astronomy and
Astrophysics and is funded by the Smithsonian Institution and the Academia
Sinica.} and Keck\footnote{Some of the data presented herein were
obtained at the W.M. Keck Observatory, which is operated as a scientific
partnership among the California Institute of Technology, the University of
California and the National Aeronautics and Space Administration.  The
Observatory was made possible by the generous financial support of the
W.~M.~Keck Foundation.} Adaptive Optics Imaging of a {\textit
Herschel}~\footnote{ {\textit Herschel} is an ESA space observatory
with science instruments provided by European-led Principal Investigator
consortia and with important participation from NASA.}-ATLAS Sub-millimeter
Galaxy at {\textit z=4.243}}

\author{
R. S. Bussmann\altaffilmark{1}, 
M. A. Gurwell\altaffilmark{1} ,
Hai~Fu\altaffilmark{2}, 
D. J. B. Smith\altaffilmark{3}, 
S. Dye\altaffilmark{4}, 
R. Auld\altaffilmark{5}, 
M. Baes\altaffilmark{6}, 
A. J. Baker\altaffilmark{7},
D. Bonfield\altaffilmark{3}, 
A. Cava\altaffilmark{8}, 
D. L. Clements\altaffilmark{9}, 
A. Cooray\altaffilmark{2}, 
K. Coppin\altaffilmark{10}, 
H. Dannerbauer\altaffilmark{11}, 
A. Dariush\altaffilmark{9}, 
G. De Zotti\altaffilmark{12}, 
L. Dunne\altaffilmark{3}, 
S. Eales\altaffilmark{4}, 
J. Fritz\altaffilmark{6}, 
R. Hopwood\altaffilmark{13}, 
E. Ibar\altaffilmark{14}, 
R. J. Ivison\altaffilmark{14}, 
M. J.  Jarvis\altaffilmark{3}, 
S. Kim\altaffilmark{2}, 
L. L. Leeuw\altaffilmark{15}, 
S. Maddox\altaffilmark{3}
M. J. Micha{\l}owski\altaffilmark{16}, 
M.  Negrello\altaffilmark{12}, 
E. Pascale\altaffilmark{17}, 
M. Pohlen\altaffilmark{17}, 
D. A. Riechers\altaffilmark{18}, 
E. Rigby\altaffilmark{19}, 
Douglas Scott\altaffilmark{20}, 
P. Temi\altaffilmark{21}, 
P. P. Van der Werf\altaffilmark{22}, 
A. Verma\altaffilmark{23},
J. Wardlow\altaffilmark{2}, 
D. Wilner\altaffilmark{1} 
 }

\altaffiltext{1}{Harvard-Smithsonian Center for Astrophysics, 60 Garden Street,
Cambridge, MA 02138, USA; rbussmann@cfa.harvard.edu}
\altaffiltext{2}{Department of Physics \& Astronomy, University of California,
Irvine, CA 92697, USA}
\altaffiltext{3}{Centre for Astrophysics, Science \& Technology Research
Institute, University of Hertfordshire, Hatfield, Herts AL10 9AB}
\altaffiltext{4}{School of Physics and Astronomy, University of Nottingham,
University Park, Nottingham NG7 2RD}
\altaffiltext{5}{Cardiff University, School of Physics \& Astronomy, Queens
Buildings, The Parade, Cardiff CF24 3AA}
\altaffiltext{6}{Sterrenkundig Observatorium, Universiteit Gent, Krijgslaan
281 S9, B-9000 Gent, Belgium}
\altaffiltext{7}{Department of Physics and Astronomy, Rutgers, the State
University of New Jersey, 136 Frelinghuysen Road, Piscataway, NJ 08854-8019,
USA}
\altaffiltext{8}{Departamento de Astrof\'isica, Facultad de CC.
F\'isicas, Universidad Complutense de Madrid, E-28040 Madrid, Spain}
\altaffiltext{9}{Imperial College London, Blackett Laboratory, Prince Consort
Road, London SW7 2AZ}
\altaffiltext{10}{Department of Physics, McGill University, Ernest Rutherford
Building, 3600 Rue University, Montreal, Quebec, H3A 2T8, Canada }
\altaffiltext{11}{Universit\"at Wien, Institut f\"ur Astronomie,
T\"urkenschanzstra\ss{}e 17, 1180 Wien, \"Osterreich}
\altaffiltext{12}{INAF - Osservatorio Astronomico di Padova;, SISSA/ISAS}
\altaffiltext{13}{Department of Physics and Astronomy, The Open University,
Walton Hall, Milton Keynes, MK7 6AA, UK}
\altaffiltext{14}{UK Astronomy Technology Centre, Royal Observatory, Blackford
Hill, Edinburgh EH9 3HJ}
\altaffiltext{15}{Department of Physics, University of Johannesburg, Auckland
Park 2006, South Africa}
\altaffiltext{16}{Scottish Universities Physics Alliance, Institute for
Astronomy, University of Edinburgh, Royal Observatory, Edinburgh, EH9 3HJ, UK}
\altaffiltext{17}{ESO, Karl-Schwarzschild-Str. 2, D-85748 Garching, Germany}
\altaffiltext{18}{Astronomy Department, California Institute of Technology, MC
249-17, 1200 East California Boulevard, Pasadena, CA 91125, USA}
\altaffiltext{19}{Institute for Astronomy, University of Edinburgh, Royal
Observatory, Edinburgh EH9 3HJ}
\altaffiltext{20}{Department of Physics and Astronomy, University of British
Columbia, Vancouver, BC V6T 1Z1, Canada}
\altaffiltext{21}{Space Science and Astrophysics Branch, NASA Ames Research
Center, MS 245-6, Moffett Field, CA 94035, USA}
\altaffiltext{22}{Leiden Observatory, Leiden University, PO Box 9513, 2300 RA
Leiden, The Netherlands}
\altaffiltext{23}{Oxford Astrophysics, Denys Wilkinson Building, University of
Oxford, Keble Road, Oxford OX1 3RH, UK}


\begin{abstract}

We present high-spatial resolution imaging obtained with the Submillimeter Array (SMA) at 880$\,\mu$m and the Keck~Adaptive Optics~(AO) system at $K_{\rm S}$-band of a gravitationally lensed sub-millimeter galaxy (SMG) at $z=4.243$ discovered in the {\it Herschel}~Astrophysical Terahertz Large Area Survey. The SMA data (angular resolution $\approx0\farcs6$) resolve the dust emission into multiple lensed images, while the Keck~AO $K_{\rm S}$-band data (angular resolution $\approx0\farcs1$) resolve the lens into a pair of galaxies separated by $0\farcs3$. We present an optical spectrum of the foreground lens obtained with the Gemini-South telescope that provides a lens redshift of~$z_{\rm lens}=0.595\pm0.005$. We develop and apply a new lens modeling technique in the visibility plane that shows that the~SMG is magnified by a factor of $\mu=4.1\pm0.2$ and has an intrinsic infrared (IR) luminosity of $L_{\rm IR} = (2.1 \pm 0.2)\times10^{13}~L_\sun$. We measure a half-light radius of the background source of $r_{\rm s} = 4.4\pm0.5~$kpc which implies an IR luminosity surface density of $\Sigma_{\rm IR}=(3.4\pm0.9)\times10^{11}~L_\sun~$kpc$^{−2}$, a value that is typical of $z>2$ SMGs but significantly lower than IR luminous galaxies at~$z\sim0$. The two lens galaxies are compact ($r_{\rm lens}\approx0.9~$kpc) early-types with Einstein radii of $\theta_{\rm E1}=0.57\pm0.01$ and $\theta_{\rm E2}=0.40\pm0.01$ that imply masses of $M_{\rm lens1}=(7.4\pm0.5)\times10^{10}~M_\sun$ and $M_{\rm lens2}=(3.7\pm0.3)\times10^{10}~M_\sun$. The two lensing galaxies are likely about to undergo a dissipationless merger, and the mass and size of the resultant system should be similar to other early-type galaxies at $z\sim0.6$.  This work highlights the importance of high spatial resolution imaging in developing models of strongly lensed galaxies discovered by {\it Herschel}.

\end{abstract}



\section{Introduction} \label{sec:intro} 

It has been known for over a decade that the star-formation rate density in the
Universe peaked around redshifts $z=1-3$ \citep[e.g.,][]{1996MNRAS.283.1388M,
1996ApJ...460L...1L}.  More recently, the advent of
bolometer arrays in the sub-millimeter (sub-mm) as well as the {\it Spitzer
Space Telescope}, have established that the contribution of
ultra-luminous infrared galaxies (ULIRGs) to the star-formation rate density in
the Universe rises sharply with redshift out to $z \sim 2$
\citep[e.g.,][]{1999MNRAS.302..632B, 2005ApJ...622..772C, 2005ApJ...632..169L,
2011ApJ...732..126M, Magnelli:2011ul}.  Although ULIRGs in the local
Universe are known to be rare \citep{1986ApJ...303L..41S} and have long been
thought to arise from a major merger of two gas-rich disk galaxies
\citep[e.g.,][]{1987AJ.....94..831A, 1996AJ....111.1025M, 1996MNRAS.279..477C,
2002ApJS..138....1B}, their nature and role in galaxy evolution at high
redshift is not yet well-understood.

The primary obstacle to studying ULIRGs at high-redshift has been one of
identification (caused in large part by faintness at optical wavelengths).
Surveys to identify ULIRGs have either been limited to small areas on the sky
\citep[e.g., the Sub-mm Common User Bolometer Array Half Degree Survey,
SHADES;][]{Coppin:2006lr}, low-spatial resolution imaging \citep[e.g.,
the Balloon-borne Large Aperture Submillimeter
Telescope;][]{2008ApJ...681..400P} or are sensitive to mid-infrared (mid-IR)
radiation which is far from the peak of the spectral energy distribution (SED)
of the ULIRG \citep[e.g.,][]{2003PASP..115..897L}.  Each of these techniques
produce samples of objects that are sufficiently faint at far-IR and sub-mm
wavelengths that follow-up observations have been time-consuming and therefore
limited to a modest number of objects, both in terms of determining redshifts
\citep[e.g.,][]{2005ApJ...622..772C} and measuring important quantities such as
accurate positions \citep[e.g.,][]{2002ApJ...573..473D, 2007ApJ...671.1531Y},
morphologies \citep[e.g.,][]{2009ApJ...693..750B, Swinbank:2010ys, 
Bussmann:2011lr}, and gas and dust masses
\citep[e.g.,][]{2005MNRAS.359.1165G, 2006ApJ...640..228T, Coppin:2008fk,
2008ApJ...680..246T, Bussmann:2009lr, 2010ApJ...712..942M,
2010ApJ...717...29K, 2011MNRAS.412.1913I, 2011ApJ...733L..11R}.

This situation is now being remedied following the launch of the {\it Herschel
Space Observatory} ({\it Herschel}$\,$).  With a large array of sensitive
detectors at 70$\,\mu$m, 100$\,\mu$m, 160$\,\mu$m, 250$\,\mu$m, 350$\,\mu$m,
and 500$\,\mu$m, the Photodetector Array Camera and Spectrometer
\citep[PACS;][]{2010A&A...518L...2P} and Spectral and Photometric Imaging
Receiver \citep[SPIRE;][]{2010A&A...518L...3G} on {\it Herschel} are
well-suited to surveying large areas of the sky at wavelengths that are ideal
for the detection of ULIRGs in the redshift range $z \sim 2-4$.  The widest
such survey is known as the {\it Herschel} Astrophysical Terahertz Large Area
Survey \citep[H-ATLAS;][]{2010PASP..122..499E} and covers 550~deg$^2$ of sky as
the largest open-time key project, reaching 5$\sigma$ sensitivities of 130~mJy
at 100$\,\mu$m, 120~mJy at 160$\,\mu$m, 32~mJy at 250$\,\mu$m, 36~mJy at
350$\,\mu$m, and 45~mJy at 500$\,\mu$m \citep{2010MNRAS.409...38I,
2011MNRAS.415..911P, 2011MNRAS.tmp..955R}.  

The wide area coverage of H-ATLAS makes it ideal for building statistically
significant samples of rare galaxies.  One such example that has been
particularly fruitful thus far is the selection of gravitationally lensed
objects.  These are systems where the light from a distant source (in this
case, a ULIRG at $z \sim 4$) is deflected by a foreground lens (typically an
early type galaxy or group of galaxies) in such a way that the background ULIRG
appears to have its angular size and brightness increased.  Several authors
have predicted that the sub-mm is an efficient waveband to identify lensing
systems due to the steep number counts of galaxies selected at sub-mm and mm
wavelengths (SMGs) \citep[e.g.,][]{1996MNRAS.283.1340B, 2002MNRAS.329..445P,
2007MNRAS.377.1557N}.  Additionally, the fact that most SMGs lie at $z>2$
\citep{2005ApJ...622..772C} increases the probability that an interloping
galaxy will lie along the line-of-sight.  Recently, \citet{Negrello:2010fk}
have shown that a selection at 500$\,\mu$m of $F_{\rm 500\mu m} > 100~$mJy
sources within the 14.4~deg$^2$ Science Demonstration Phase field of H-ATLAS
identifies strongly lensed systems, low-$z$ spiral galaxies
\citep{2005MNRAS.356..192S}, and higher-$z$ active galactic nuclei (AGN) that
are radio-bright and show a synchrotron emission spectrum even into the SPIRE
bands \citep{2005A&A...431..893D}.  Shallow ground-based optical and radio
imaging can be used to remove the latter two classes of objects, leaving only
the strongly lensed systems.

The H-ATLAS source catalog already extends to $\approx 130~$deg$^2$ (Phase~1
catalog; Dunne et al., in prep.).  This paper focuses on one source of
particular interest drawn from the Phase~1 catalog: H-ATLAS~J142413.9+022304
\citep[this object is denoted ``ID~141'' in][hereafter, we refer to it as
G15v2.779]{Cox:2011fk}.  SPIRE photometry of this source shows that it is one
of the brightest detected so far in {\it Herschel} wide-field surveys and that
its SED peaks at wavelengths greater than 500$\,\mu$m, suggesting that it lies
at $z > 3$.  This source has been the target of significant follow-up efforts:
the Plateau de Bure Interferometer (PdBI) has detected millimeter (mm) and
sub-mm CO emission lines which imply that the redshift of this source is
$z=4.243 \pm 0.001$, while data from the Atacama Pathfinder Experiment (APEX)
have shown that the dominant cooling line in this galaxy is [CII] emission
\citep{Cox:2011fk}.  This makes this one of the few SMGs known at $z > 4$
\citep[][]{Capak:2008lr, Schinnerer:2008uq, Coppin:2009lr, Daddi:2009kx,
Daddi:2009qy}.  In addition, a faint counterpart ($r=22.06$ AB) is detected in
both the Sloan Digital Sky Survey \citep[SDSS DR7,][]{2000AJ....120.1579Y} and
the United Kingdom Infrared Deep Sky Survey
\citep[UKIDSS,][]{2007MNRAS.379.1599L} that has a photometric redshift of
$z_{\rm lens} = 0.69 \pm 0.13$ \citep{2011MNRAS.416..857S}.  Altogether, the
evidence favors a scenario in which the background source is a SMG at high
redshift that is being gravitationally lensed by an object at intermediate
redshift, consistent with the lensing hypothesis of \citet{2007MNRAS.377.1557N,
Negrello:2010fk}.  

Recent observations by the Submillimeter Array (SMA) have shown an elongation
along the southeast---northwest direction, possibly an indication of
interesting morphological features on scales smaller than $\approx 2\arcsec$
\citep{Cox:2011fk}.  In this paper, we present high-spatial resolution
($0\farcs6$) SMA imaging at 880$\,\mu$m, Keck Adaptive Optics (AO) $K_{\rm
S}$-band imaging, and Gemini Multi-Object Spectrograph-South (GMOS-S) optical
spectroscopy of this object and demonstrate their utility for constraining a
detailed model of the lens-source system.  We use the SMA, Keck, and Gemini
data to probe the source size and magnification factor as well as the luminous
plus dark matter mass of the lensing galaxies.  The magnification factor is a
critical parameter, since it is needed to understand intrinsic properties of
the background SMG such as its IR luminosity ($L_{\rm IR}$, integrated over
8-1000$\,\mu$m) as well as molecular gas and dust masses ($M_{\rm gas}$ and
$M_{\rm dust}$).  G15v2.779 is an example of a gravitationally lensed system
discovered in H-ATLAS that permits the simultaneous study of obscured
star-formation at high redshift as well as the nature of light and dark matter
in galaxies at intermediate redshift.

When the H-ATLAS catalogs are complete, we expect to have $\approx 300$
candidate lensed systems.  In addition, systems identified from the {\it
Herschel} Multi-tiered Extragalactic Survey
\citep[HerMES;][]{2010A&A...518L..21O} could bring this total to 500 such
objects discovered by {\it Herschel}.  Recent efforts using near-IR imaging to
push to fainter sub-mm flux densities and grow the list of candidates up to
$\sim 2000$ appear promising \citep{Gonzalez-Nuevo:2012lr}.  Using similar
selection techniques, the South Pole Telescope (SPT) has identified a sample of
$\approx 40$ candidate lensed SMGs within the initial 87~deg$^2$ survey
\citep{Vieira:2010rr}.  The final survey area will cover 2000~deg$^2$ and is
expected to provide a sample of $\sim 1000$ lensed SMGs.  Due to the selection
at 1.4mm, the SPT sample of lensed SMGs will be complementary to the {\it
Herschel} sample in the sense that it will be biased towards higher redshift or
cooler dust temperatures.  Together, both the {\it Herschel} and SPT samples
offer the opportunity to expand upon the legacy of work on strongly lensed
galaxies over a similar redshift range undertaken as part of the Center for
Astrophysics Arizona Space Telescope Lens Survey
\citep[CASTLeS;][]{Munoz:1998mz}, the Cosmic Lens All-Sky Survey
\citep[CLASS;][]{Myers:2003lr, Browne:2003lr}, and the Jodrell Bank Very Large
Array gravitational lens survey \citep[JVAS;][]{King:1996fk} by increasing the
sample size of such systems by 1-2 orders of magnitude.  In comparison to
lensed systems selected via optical spectroscopy \citep[e.g., the Sloan Lens
Advanced Camera for Surveys or SLACS and the Baryon Acoustic Oscillation Survey
Emission-Line Lensing Survey or BELLS;][]{Bolton:2008wd, Brownstein:2012cr},
the (sub-)mm selection is highly complementary in that it identifies (1) lensed
galaxies that are both more luminous and at higher redshift; and (2) lensing
galaxies that span a wider range in optical brightness (in particular, they do
not need to be bright enough to be detected in SDSS-III spectroscopy).

Throughout this paper we assume $H_0=$71~km~s$^{-1}$~Mpc$^{-1}$, $\Omega_{\rm
m} = 0.27$, and $\Omega_\lambda = 0.73$.  At $z=4.243$, this results in a scaling
of 6.9~kpc~arcsec$^{-1}$.

\section{Observations}\label{sec:obs}

\subsection{SMA Data}\label{sec:smaobs}

SMA imaging data of G15v2.779 was initially obtained as a short observation
conducted in the compact array configuration on 2010 June 16 \citep[$t_{\rm
int} = 3$~hrs on-source integration time, see][]{Cox:2011fk}.  These
data yielded a robust detection of the target and provided a total flux density
measurement at 880$\,\mu$m of $F_{\rm 880\mu m} = (90 \pm 2)$~mJy.  The data
also pinpointed the location of the {\it Herschel} source: RA 14:24:13.98, Dec
+02:23:03.45 (J2000.0) and with a beamsize of 2$\farcs4 \times 1\farcs3$ hinted
at an elongation in the northwest---southeast direction \citep[$-32\pm4$
degrees east of north;][]{Cox:2011fk}.  

Subsequent data in the very extended array configuration were obtained on 2011
January 4 and 6 ($t_{\rm int} = 3$~hrs in total, max baseline length of 475~m).
Extended array configuration data were obtained to improve {\it uv} coverage on
2011 January 28 ($t_{\rm int} = 2$~hrs, max baseline length of 226~m).
Atmospheric opacity was low ($\tau_{\rm 225 GHz} < 0.08$) and phase stability
was good (phase errors less than 30$^\circ$).

We optimized the SMA single-polarization 345~GHz receivers for continuum
detection by tuning the primary local oscillator to 339.58 GHz and an
intermediate frequency coverage of 4--8 GHz, providing a total of 8~GHz
bandwidth (considering both sidebands).  The observations did not cover the
nearest CO emission line, CO($J=16-15$) at about 351~GHz since that frequency
is near the high end of the range of the SMA 345~GHz receivers and would have
compromised the sensitivity of our continuum observations.

We used the Interactive Data Language (IDL) {\sc MIR} package to calibrate the
{\it uv} visibilities.  The blazar 3C279 was used as the primary bandpass
calibrator while Titan was used as the absolute flux calibrator.  The nearby
quasars 1337$-$129 ($F_{\rm 880\mu m} = 1.7~$Jy, 19 degrees from target) and
1512$-$090 ($F_{\rm 880\mu m} = 1.5~$Jy, 17 degrees from target) were used for
phase gain calibration.  The quasar 1458+042 ($F_{\rm 880\mu m} = 0.15~$Jy, 9
degrees from target) was observed to provide an independent check of the
reliability of the calibration, particularly phase transfer.  

For imaging, we used the Multichannel Image Reconstruction, Image Analysis, and
Display (MIRIAD) software package \citep{1995ASPC...77..433S} with natural
weighting for maximum sensitivity.  This resulted in an elliptical gaussian
beam with a full-width half-maximum (FWHM) of $0\farcs69\times0\farcs50$ and
position angle of 62.4 degrees east of north.

\begin{figure}[!tbp] 
\begin{centering}
\includegraphics[width=0.45\textwidth]{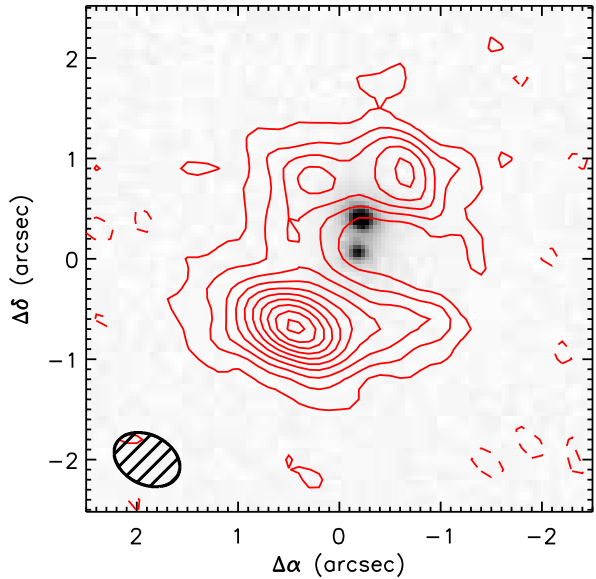}

\caption{ Multi-wavelength imaging of G15v2.779 centered on RA~14:24:13.975,
Dec~+02:23:03.60 (the 880$\mu$m emission centroid).  Red contours highlight SMA
880$\,\mu$m dust continuum emission from the lensed SMG at $z=4.243$ (drawn
at -2, 2, 4, 6, ... times the 1$\sigma$ rms level, where $\sigma = 1.0$~mJy).
The size of the SMA beam (FWHM $0\farcs69 \times 0\farcs50$) is shown by the
black hatched ellipse at the bottom left corner of the plot.  The greyscale
background shows Keck AO $K_{\rm S}$-band imaging which has resolved the lens
into two early type galaxies.  \label{fig:imaging}}
\end{centering}

\end{figure}

Figure~\ref{fig:imaging} shows the SMA image of this source (combining compact,
extended and very extended array configurations) in red contours.  The sub-mm
emission is resolved into two bright emission regions to the southeast and
northwest of the lensing galaxies.  These emission regions have peak
intensities of $S_{\rm 880\mu m} = (21 \pm 1.0)$~mJy~beam$^{-1}$ and $S_{\rm
880\mu m} = (10 \pm 1.0)$~mJy~beam$^{-1}$, respectively (quoted 1$\sigma$
uncertainties do not include the estimated absolute calibration uncertainty of
10\%). The peaks are located at positions of (RA=14:24:14.006,
Dec=+02:23:02.81) and (RA=14:24:13.938, Dec=+02:23:04.40), respectively (the
uncertainty in the relative position of these peaks is $\approx 0\farcs05$).
In addition, there is a background of complex substructure that, together with
the two bright emission regions, sums to a total flux density of $F_{\rm 880\mu
m} = 90 \pm 1.0$~mJy and is likely to be produced by gravitational lensing.  

\subsection{Keck Adaptive Optics Data}\label{sec:keckobs}

We obtained a 1920~sec $K_{\rm S}$-band image of G15v2.779 on 2011 April 13
(UT) as part of program ID C213N2L \citep[PI: H. Fu; e.g.,][]{Fu:2012uq} with the
Keck~II laser guide-star adaptive-optics system
\citep[LGSAO;][]{2006PASP..118..297W}. An $R=17.9$ magnitude star 26$\arcsec$
south of G15v2.779 served as the tip-tilt reference star. The expected Strehl
ratio is about 0.17 at the source position.  We used the second generation
near-infrared camera
(NIRC2\footnote{http://www2.keck.hawaii.edu/inst/nirc2/Manual/ObserversManual.html})
at $0\farcs04$~pixel$^{-1}$ scale and executed a nine-point dithering pattern
with 3$\arcsec$ dithering steps. Three 80~sec exposures were obtained at each
dithering position.  The natural seeing was about $0\farcs7$.

We used our own IDL program to reduce the images.  After dark subtraction and
flat-fielding, sky background and object masks were updated iteratively. For
each frame, after subtracting a scaled median sky, the residual background was
removed with B-spline models. In the last iteration, we discarded the three
frames of the poorest image quality and corrected the NIRC2 geometric
distortion using the solution of
P.~B.~Cameron\footnote{http://www2.keck.hawaii.edu/inst/nirc2/forReDoc/post\_observing/dewarp/}
before combining the aligned frames. The resolution of the final image is
$0\farcs1$ in FWHM, as measured from the stellar source $4\farcs5$ northwest of
G15v2.779. Astrometry was determined from SDSS photometric sources inside the
40$\arcsec$ field of view and carries 1$\sigma$ uncertainties in an absolute
sense of $\approx 0\farcs4$ (in fact, we show in section~\ref{sec:results} that
the tightest constraints on the astrometry are derived directly from the lens
modeling).  The flux scale in the image was normalized such that the total flux
of the two lensing galaxies matches that seen in the UKIDSS $K$-band data,
where the two galaxies have a total magnitude of $K = 17.89 \pm 0.17$.

\begin{figure}[!tbp] 
\begin{centering}
\includegraphics[width=0.45\textwidth]{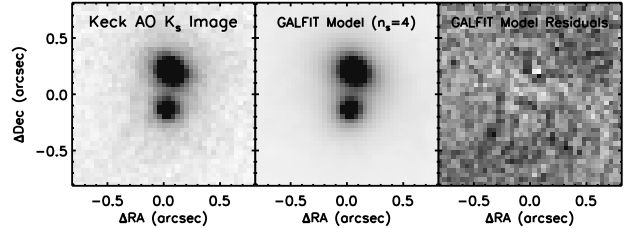}

\caption{ GALFIT modeling of G15v2.779.  {\it Left}: Keck AO $K_{\rm S}$-band
image (as in Figure~\ref{fig:imaging}).  {\it Middle}: Best-fit GALFIT model
(assuming $n_s = 4$ for both lensing galaxies).  {\it Right}: Residuals
obtained after subtracting the best-fit model from the Keck data.  Both
morphologies are consistent with early-type galaxies.  \label{fig:GALFIT}}
\end{centering}

\end{figure}

The grayscale of Figure~\ref{fig:imaging} shows the Keck AO $K_{\rm S}$-band
image of G15v2.779.  The backgound SMG is undetected, while the two
foreground lensing galaxies are detected at high significance.  The AO imaging
indicates the secondary lens galaxy is located 0$\farcs$025 east and
0$\farcs$327 south of the primary.  A {\sc Galfit} decomposition of the two
sources seen in the Keck image into S\'ersic components indicates that a de
Vaucouleurs profile is appropriate for both the northern and southern galaxies.
In fact, the best-fit models have $n_{\rm s} > 4$, but this is probably due to
faint, large-scale fluctuations in the background sky level -- hereafter, we
assume $n_{\rm s} = 4$ for both galaxies for simplicity.  
Figure~\ref{fig:GALFIT} shows the best-fit {\sc Galfit} model and the residuals after
subtracting the model from the data.  Table~\ref{tab:galfitresults} contains
the best-fit parameters from the {\sc Galfit} modeling and their 1$\sigma$
uncertainties.  These are underestimates of the true errors as they do not
account for degeneracies between the parameters.  The two lensing galaxies are
highly compact, with effective radii of $r_{\rm lens1} = 0.84\pm0.01$~kpc and
$r_{\rm lens2} = 0.91 \pm 0.03$~kpc.  The northern galaxy has $K_{\rm S} =
18.22 \pm 0.17$ (Vega mag) and the southern galaxy has $K_{\rm S} = 19.01 \pm
0.17$ (Vega mag).  

The background source is not detected inside a $1\farcs5$ radius, implying a
5$\sigma$ limit of $K_{\rm S} > 20.14$ (Vega mag).  This corresponds to a flux
density 5$\sigma$ limit of $F_{K_{\rm S}} < 5.8\;\mu$Jy.  This level of
faintness in the near-IR is frequently an indication of $z > 4$ systems
\citep[e.g.][]{2002ApJ...573..473D, Daddi:2009kx}.  If differential
magnification effects are insignificant (i.e., $\mu_{K_{\rm S}} \approx
\mu_{\rm 880\mu m}$), then this implies a 500$\mu$m to $K_{\rm S}$-band flux
density ratio of $\approx 35000$.  At $z=4.243$, this corresponds roughly to
rest-frame 100$\mu$m to $U$-band.  An interesting comparison example is
Mrk~231, a heavily obscured ULIRG in the local universe that has a 100$\mu$m to
$U$-band flux density ratio of $\approx 10000$.  This value is about a factor
of 3-4 lower than G15v2.779, suggesting that the obscuration in G15v2.779 is
extreme.  In section~\ref{sec:bg}, we will compare the obscuration in G15v2.779
with other $z \sim 4$ SMGs.

\begin{deluxetable}{lccc}
\tabletypesize{\small} 
\tablecolumns{3}
\tablewidth{0pt}
\tablecaption{GALFIT Lens Modeling Results\tablenotemark{a}}
\tablehead{
\colhead{} & \colhead{Lens 1} & \colhead{Lens 2}
}
\startdata
$m_{K_S}$ (Vega) & $18.22\pm0.01$ & $18.98\pm0.02$ \\
$n_{\rm lens}$\tablenotemark{b} & 4 & 4   \\
$r_{\rm lens}$ (kpc) & $0.84\pm0.01$ & $0.91\pm0.03$  \\
$\epsilon_{\rm lens}$ & $0.29\pm0.01$ & $0.06\pm0.02$   \\
$\phi_{\rm lens}$ (deg) & $62\pm1$ & $-47\pm14$   \\
\tablenotetext{a}{Uncertainties do not reflect degeneracies between the
parameters or absolute calibration uncertainty in $m_{K_S}$.}
\tablenotetext{b}{S\'ersic indices are fixed to be $n_{\rm lens} = 4$ (see text
for details).}
\enddata
\label{tab:galfitresults}
\end{deluxetable}

\subsection{Gemini GMOS-S Optical Spectroscopy}\label{sec:gmos}

Long-slit spectroscopic observations of G15v2.779 were taken using the Gemini
GMOS-South instrument on the night of 2011 March 6, under photometric
conditions as part of programme GS-2011A-Q-57 (PI: D.~J.~B.~Smith). Two
observations of 1800 seconds each were made through a 2$\arcsec$ slit, using
the R400 grating and the OG515 blocking filter, with dithering both in the
wavelength direction and along the slit to minimize the effects of bad columns
and gaps between the GMOS-S chips.  The central wavelengths for the two
observations were 630 and 635 nm, and flat field observations were interspersed
between the observations at each wavelength setting, as recommended by the
Gemini observatory.  CuAr arc lamp exposures were taken for the purposes of
wavelength calibration, using the same instrumental setup as for the science
exposures, and the spectral resolution obtained was $\approx$6.0\AA. A position
angle of 330$^\circ$ East of North was chosen to place the slit along the major
axis of the extension seen in the SMA and PdBI observations, and the CCD was
binned by 4 pixels in both the spectral and spatial directions. The long-slit
data were reduced using the {\sc IRAF} Gemini GMOS reduction routines, and
following the standard GMOS-S reduction steps in the example taken from the
Gemini observatory webpages. Since the primary aim of these observations was to
obtain a spectroscopic redshift and measure of the Balmer/4000 \AA\ break for
the lensing source (or sources), flux calibration was not performed.

\begin{figure}[!tbp] 
\begin{centering}
\includegraphics[width=0.45\textwidth]{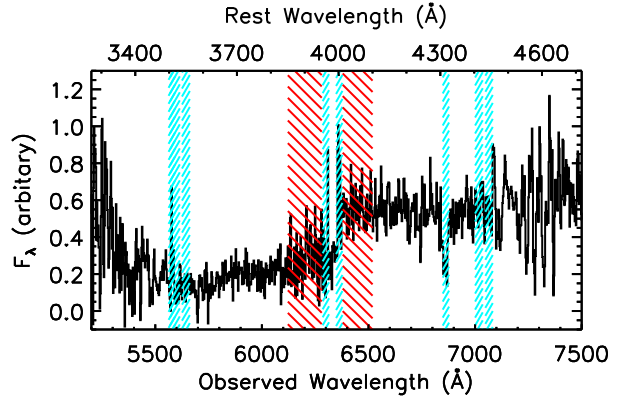}

\caption{ Gemini GMOS-S optical spectrum of G15v2.779.  Narrow cyan hatched
lines indicate regions of the spectrum corrupted by either chip gaps in the CCD
or atmospheric spectral features.  Broad red-hatched regions indicate the
portions of the spectrum used to compute $D_n (4000)$ and are selected to avoid
the cyan regions.  The background source is hidden by dust at these
wavelengths, but both foreground lens galaxies lie within the 2$\arcsec$ GMOS-S
slit.  The spectrum shows a strong, abrupt break at an observed-frame
wavelength of $\lambda_{\rm obs} = 6350\,$\AA, likely corresponding to the
Balmer/4000$\,$\AA\ break and implying a lens redshift of $z_{\rm lens} =
0.595\pm0.005$.  The strength of the break implies the lensing galaxies are
dominated by old stars \citep[5-10~Gyr, depending on
metallicity;][]{Kauffmann:2003qf}. \label{fig:gmos}}
\end{centering}

\end{figure}

Figure~\ref{fig:gmos} shows the optical spectrum of G15v2.779 obtained with
Gemini GMOS-S.  Because the slit width is 2$\arcsec$, both lensing galaxies are
included in this spectrum.  A strong break at observed-frame 6350 \AA\ is
obvious in the spectrum.  Two narrow emission lines on either side of this
break coincide with atmospheric features and are likely not produced by
astronomical sources.  The break in the spectrum likely corresponds to a
Balmer/4000 \AA\ break and suggests that both lensing galaxies are at
$z=0.595\pm0.005$ \citep[the redshift and error were estimated using a by-eye
comparison of the observed spectrum with a synthesized simple stellar
population with solar metallicity and an age of 5~Gyr;][]{2003MNRAS.344.1000B}.
We note that without spatially resolved spectroscopy it is not possible to
confirm that both lensing galaxies are located at the same redshift.  The ratio
of the flux density longward and shortward of the break is known as the
$D_n\;(4000)$ value and is a measure of the average age of the stars within the
galaxies.  We use the definition from \citet{Kauffmann:2003qf} and measure
$D_n\;(4000) = 2.0\pm0.2$ (we adjust the wavelength window longward of the
break to avoid the portion of the spectrum affected by atmospheric emission
lines).  This value is typical of galaxies dominated by old stellar populations
\citep[5-10~Gyr, depending on metallicity;][]{Kauffmann:2003qf}.

\section{A Detailed Lens Model}\label{sec:results}

The combination of good sensitivity and high spatial resolution provided by the
SMA and Keck AO data (probing the emission from the lensed and lensing
galaxies, respectively) permits a detailed study of the parameters of the
lensing model that describe this system.  The lensed emission is detected only
in the SMA data.  Since the SMA is an interferometer, the surface brightness
map of the lensed emission is obtained with incomplete {\it uv} coverage,
implying that surface brightness is not necessarily conserved and that the
pixel-to-pixel errors in the map are correlated.  Furthermore, the lensed
emission observed by the SMA comprises multiple, resolved components.  For
these reasons, it is important to compare model and data in the visibility
plane rather than the image plane.  


We make use of the publicly available {\sc Gravlens} software
\citep{2001astro.ph..2340K} to map emission in the source plane to the image
plane for a given lensing mass distribution.  Using this software, a model of
the lensed emission is constructed based on the density profile of the lens --
assumed here to be two singular isothermal ellipsoids (one for each galaxy seen
in the Keck AO imaging), the morphology of the source -- taken here to be a
S\'ersic profile, and the position of the source relative to the lens.
Although the S\'ersic profile represents a crude simplification of the true
background source morphology, we use it here because it permits a test of the
variation in the lensing properties (e.g., magnification of the background
source) as a function of variation in the nature of the source (e.g.,
half-light radius).  Later in this section, we discuss how our best-fit
parameters change when a second source is added to the source plane.  

The {\sc Gravlens} lens model (with a single source in the source plane)
contains 15 free parameters: the position of the source relative to the SMA
880$\mu$m emission centroid ($\Delta \alpha_{\rm s}$ and $\Delta \delta_{\rm
s}$), the intrinsic flux of the source ($F_{\rm s}$), the half-light radius of
the source ($r_{\rm s}$), the S\'ersic index of the source ($n_{\rm s}$), the
ellipticity (defined as $\frac{a-b}{a}$) and position angle of the source
($\epsilon_{\rm s}$ and $\phi_{\rm s}$), the position of the primary lens
relative to the SMA emission centroid ($\Delta \alpha_{\rm lens1}$ and $\Delta
\delta_{\rm lens1}$), the mass of the primary lens (parameterized in terms of
the angular Einstein radius, $\theta_{\rm E1}$), the ellipticity and position
angle of the primary lens ($\epsilon_{\rm lens1}$ and $\phi_{\rm lens1}$), and
the Einstein radius, ellipticity, and position angle of the secondary lens
($\theta_{\rm E2}$, $\epsilon_{\rm lens2}$, and $\phi_{\rm lens2}$).  

The 1-$\sigma$ absolute astrometric accuracy between the SMA and Keck images is
$0\farcs4$, so in our modeling efforts we allow the position of the lens to
vary by as much as $0\farcs8$ in both RA and Dec (i.e., 2$\sigma$ in each
direction).  In fact, the constraints from the lens model are significantly
tighter than $0\farcs4$, as we show below.  

We make use of the Keck AO $K_{\rm S}$-band data and fix the position of the
secondary lens relative to the primary at $\Delta \alpha_{\rm lens2} =
0\farcs025$ and $\Delta \delta_{\rm lens2} = -0\farcs327$ (rotational
astrometric uncertainties between the SMA and Keck AO images are sub-dominant
to translational uncertainties).  Furthermore, we use our {\sc GALFIT} results
and constrain the ellipticity and position angle of the lenses to be within
3$\sigma$ of the best-fit {\sc Galfit} values (as given in
Table~\ref{tab:galfitresults}.  We tested models in which the mass of each
foreground lensing galaxy was allowed to vary as well as models in which the
mass of the secondary lens was fixed to be equal to one-half that of the
primary lens.  This latter choice assumes that both lens
galaxies are located at the same redshift $z_{\rm lens}$ and is supported by
the ground-based Gemini GMOS-S spectrum of G15v2.779 (see
section~\ref{sec:gmos}).  High-spatial resolution optical or near-IR
spectroscopy are needed to resolve the two lensing galaxies and provide a
definitive test of the validity of our assumptions.

For a given set of model parameters, {\sc Gravlens} generates a surface
brightness map of the lensed emission.  This surface brightness map can then be
used as input to MIRIAD's {\sc uvmodel} task, which computes the Fourier
transform of the image and samples the resulting visibilities in a way that
matches the sampling of the actual observed SMA visibility dataset ($V_{\rm
SMA}$) to produce a ``simulated visibility'' dataset ($V_{\rm model}$).  The
quality of fit for a given set of model parameters is determined from the
chi-squared value ($\chi^2$) according to the following equation:

\begin{equation}
    \chi^2 = \sum_{u, v}^N \frac{(V_{\rm SMA}(u,
    v) - V_{\rm model}(u, v))^2}{\sigma(u, v)^2},
\end{equation}

\noindent where $\sigma(u, v)$ is the 1$\sigma$ uncertainty level for each
visibility and is determined from the system temperatures (this corresponds to
a natural weighting scheme).  Because the measured visibilities are complex, we
compute both $\chi_{\rm real}^2$ and $\chi_{\rm imag}^2$ and measure the total
$\chi^2$ as the sum of the real and imaginary components.

We employ a Markov chain Monte Carlo (MCMC) technique to sample the posterior
probability density function (PDF) of our model parameters.  In particular, we
use the {\sc emcee} code \citep{Foreman-Mackey:2012lr} to implement the MCMC analysis.
The algorithm adopted by {\sc emcee} was originally presented in
\citet{goodmanweare} and uses an affine-invariant ensemble sampler to
obtain significant performance advantages over standard MCMC sampling methods.  

Here, we summarize the behavior of the ensemble sampler technique.  The
available parameter space is searched using a set of $N_{\rm walkers}$ walkers.
During iterations of the MCMC, each walker selects another walker from the
ensemble and identifies a new position in parameter space based on the
positions of both walkers (this is known as a ``stretch move'').  Once a new
position has been found, the posterior PDF is computed and compared to that of
the previous position.  New positions with higher probability (i.e., lower
$\chi^2$) are always accepted; those with lower probability are {\it sometimes}
accepted.  After a sufficient number of iterations ($N_{\rm iter}$), the ensemble
of walkers samples the parameter space in a way that reflects the posterior
PDF.  The mean and variance of each parameter can then be measured directly
from the history of walker positions.  We employ a ``burn-in'' phase with
$N_{\rm walkers} = 250$ and $N_{\rm iter} = 200$ (i.e., 50,000 samplings of the
posterior PDF) that is used to identify the best-fit model position.  This
position is then used to initialize the ``final'' phase with $N_{\rm walkers} =
250$ and $N_{\rm iter} = 40$ (i.e., 10,000 samplings of the posterior PDF).  To
ensure that the posterior PDF was sampled with a sufficient number of
iterations, we computed the autocorrelation time for each parameter in a given
ensemble of walkers and found that it was of order unity for each parameter.
This implies that we have 10,000 independent samplings of the posterior PDF,
more than enough to obtain a robust measurement of the mean and uncertainty
on each parameter of the model.

During each MCMC iteration, we also measure the magnification factor $\mu$ (we
follow the nomenclature in the SMG literature here and use $\mu$ to refer to
the total magnification obtained by summing over all individual lensed
components) using the following method.  First, we take the unlensed, intrinsic
source model and measure the total flux density ($F_{\rm in}$) within an
elliptical aperture ($A_{\rm in}$) centered on the source with ellipticity and
position angle equal to that of the source model and with a semi-major axis
length of $5\arcsec$.  Next, we take the lensed image of the best-fit model and
measure the total flux density ($F_{\rm out}$) within the aperture $A_{\rm
out}$, where $A_{\rm out}$ is determined by using {\sc Gravlens} to map $A_{\rm
in}$ in the source plane to $A_{\rm out}$ in the image plane (using the lens
parameters which correspond to the best-fit model).  The magnification is then
computed simply as $\mu = F_{\rm out} / F_{\rm in}$.  The best-fit value and
1$\sigma$ uncertainty are drawn from the posterior PDF as with the other
parameters of the model.

\begin{figure*}[!tbp] 
\epsscale{1.00} 
\includegraphics[width=\textwidth]{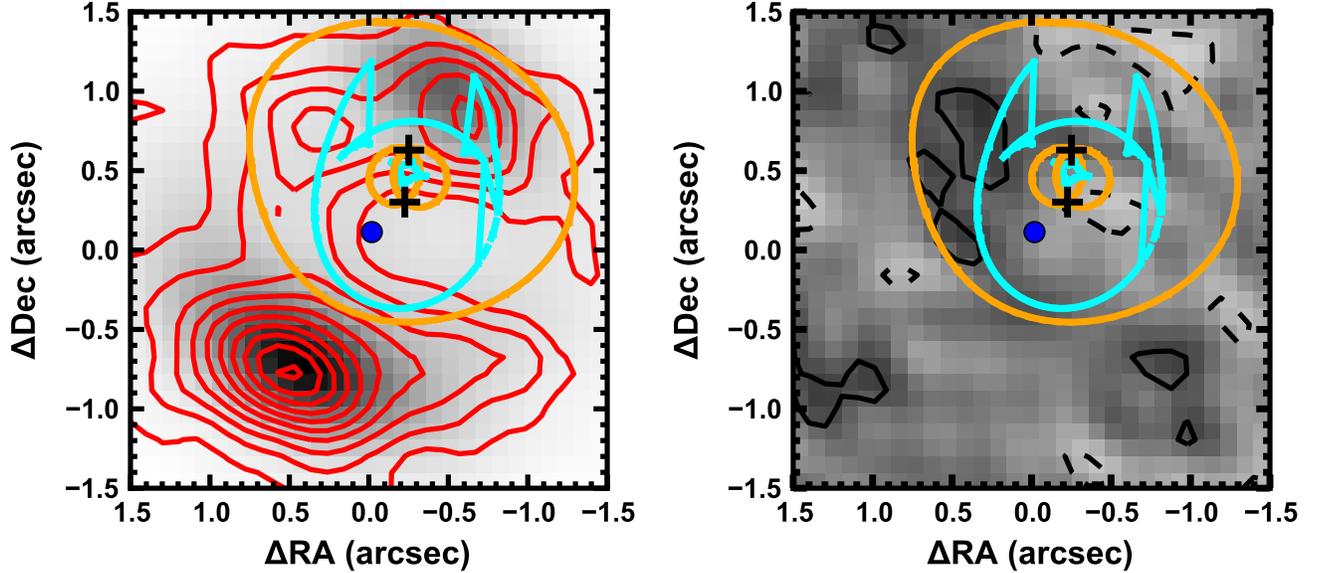}

\caption{ Comparison of best-fit {\sc Gravlens} model with SMA data.  {\it
Left}: SMA imaging (red contours) overlaid on the inverted, deconvolved map of
the best-fit model visibilities.  {\it Right}: Residual image obtained by
inverting and deconvolving the residual visibilities (i.e., the cleaned map of
the difference between the model and data visibilities).  Critical curves and
caustics are traced by orange and cyan lines, respectively.  Lens positions as
derived from the Keck AO $K_{\rm S}$-band imaging are marked by plus signs.
The peak flux position of the source is shown with a filled blue circle.
Contours indicate -2, 2, 4, 6, etc. times the 1$\sigma$ rms noise level.  This
model has a magnification factor of $\mu = 4.1 \pm 0.2$ and Einstein radii of
$\theta_{\rm E1} = 0\farcs57 \pm 0\farcs01$ and $\theta_{\rm E1} = 0\farcs40
\pm 0\farcs01$.  \label{fig:modeling}}

\end{figure*}

The best-fit model (assuming a fixed mass ratio between the primary and
secondary lens) produced by {\sc Gravlens} is shown in
Figure~\ref{fig:modeling} and demonstrates that many of the features present in
the SMA imaging can be reproduced in detail by {\sc Gravlens}.  The panel on
the left shows the SMA imaging overlaid on the inverted, deconvolved map of the
best-fit model visibilities. The panel on the right shows the residual image
obtained by inverting and deconvolving the residual visibilities (i.e., the
cleaned map of the difference between the model and data visibilities).  The
model fits the two brightest components in the SMA image (the peaks to the
southeast and northwest), but fails to reproduce fully the peaks in the map to
the northeast and southwest.

We measure the following parameters of interest from the model:  $\mu =
4.1\pm0.2$, $\theta_{\rm E1} = 0\farcs57\pm 0\farcs01$, $\theta_{\rm E2} =
0\farcs40\pm 0\farcs01$, $\Delta \alpha_{\rm lens1} = -0\farcs27 \pm
0\farcs03$, $\Delta \delta_{\rm lens1} = 0\farcs63 \pm 0\farcs03$ (this
position is within 1$\sigma$ of the position indicated from the Keck AO
astrometry), $\Delta \alpha_{\rm s} = 0\farcs03 \pm 0\farcs02$, $\Delta
\delta_{\rm s} = 0\farcs10 \pm 0\farcs02$, $n_{\rm s} = 2.9\pm0.3$, $r_{\rm s}
= 4.4 \pm 0.5$~kpc, $\epsilon_{\rm s} = 0.27 \pm 0.09$, and $\phi_{\rm s} = 77
\pm 12$ degrees east of north.  The best-fit model has $\chi^2 = 217799.1$ and
144191 degrees of freedom.  These constraints on model parameters are reported
in Table~\ref{tab:lensingresults}.  

Also reported in Table~\ref{tab:lensingresults} are the results obtained when
the mass of the secondary is allowed to vary relative to the primary.  The
uncertainties on the mass of each individual lensing galaxy are significantly
larger in this case because our lens model constrains the sum of the masses of
the two lensing galaxies.  The constraint on the sum of the masses is
$\theta_{\rm E1}+\theta_{\rm E2} = 0\farcs97\pm 0\farcs02$ (consistent with the
results from models where the mass ratio has been fixed).  The Gemini GMOS-S
optical spectrum provides evidence that the two lensing galaxies are both
$z=0.595$.  Under that assumption, the secondary is significantly less luminous
and hence likely to be significantly less massive.  Therefore, for the
remainder of the paper we have assumed a 2:1 mass ratio between the primary and
secondary.  Spatially resolved spectroscopy of the lensing galaxies is needed
to prove the validity of this assumption.

\begin{deluxetable}{lcccc}
\tabletypesize{\small} 
\tablecolumns{5}
\tablewidth{250pt}
\tablecaption{Gravitational Lens Model Results}
\tablehead{
\colhead{} & 
\colhead{Single-source} & \colhead{Single-source} & \colhead{Two-source} \\
\colhead{} & \colhead{$\theta_{\rm E1}/\theta_{\rm E2}$ free} & \colhead{$\theta_{\rm
E1}/\theta_{\rm E2}$ fixed} & \colhead{$\theta_{\rm
E1}/\theta_{\rm E2}$ fixed}
}
\startdata
$\Delta \alpha_{\rm s1}$ ($\arcsec$)     & $0.04\pm0.02$     & $0.03\pm0.03$  & $-0.05\pm0.02$ \\ 
$\Delta \delta_{\rm s1}$ ($\arcsec$)     & $0.09\pm0.02$     & $0.10\pm0.02$  & $ 0.09\pm0.03$ \\ 
$r_{\rm s1}$ (kpc)                       & $3.2\pm0.5$       & $4.4\pm0.5$    & $3.9\pm0.6$  \\   
$n_{\rm s1}$                             & $2.9\pm0.3$       & $2.9\pm0.3$    & $2.8\pm0.5$   \\             
$\epsilon_{\rm s1}$                      & $0.24\pm0.09$     & $0.27\pm0.09$  & $0.38\pm0.09$  \\   
$\phi_{\rm s1}$ (deg)                    & $84\pm16$         & $77\pm12$      & $80\pm10$  \\     
$\Delta \alpha_{\rm s2}$  ($\arcsec$)    & ---               & ---            & $-0.2\pm0.1$ \\  
$\Delta \delta_{\rm s2}$  ($\arcsec$)    & ---               & ---            & $0.53\pm0.09$ \\  
$r_{\rm s2}$  (kpc)                      & ---               & ---            & $2.2\pm0.8$  \\   
$n_{\rm s2}$                             & ---               & ---            & 0.5\tnm{a}  \\             
$\epsilon_{\rm s2}$                      & ---               & ---            & $0.3\pm0.1$  \\   
$\phi_{\rm s2}$ (deg)                    & ---               & ---            & $90\pm30$  \\     
$\Delta \alpha_{\rm lens1}$ ($\arcsec$)  & $-0.28\pm0.03$    & $-0.27\pm0.03$ & $-0.32\pm0.04$ \\ 
$\Delta \delta_{\rm lens1}$ ($\arcsec$)  & $0.65\pm0.04$     & $0.63\pm0.03$  & $0.53\pm0.04$ \\ 
$\theta_{\rm E1}$ ($\arcsec$)            & $0.44\pm0.09$     & $0.57\pm0.01$  & $0.57\pm0.02$ \\  
$\theta_{\rm E2}$ ($\arcsec$)            & $0.53\pm0.09$     & $\theta_{\rm E1} / \sqrt{2}$  & $\theta_{\rm E1} / \sqrt{2}$   \\ 
$\mu$                                    & $4.2\pm0.3$       & $4.1\pm0.2$    &  $3.7\pm0.3$ \\   
$\chi^2$                & 217796.9           & 217799.1           &  217769.2 \\          
$N_{\rm DOF}$           &  144190            &  144191            &  144185 \\
\tnt{a}{~S\'ersic index of secondary source is fixed at $n_{\rm s2} = 0.5$.
See section~\ref{sec:results} for details.}
\enddata
\label{tab:lensingresults}
\end{deluxetable}

The residual map in Figure~\ref{fig:modeling} shows that our single-source
model fails to account for the secondary peaks in the map to the northeast and
southwest as well as a faint ring of emission partially lined up with the
tangential critical curve.  We investigated whether adding an additional source
in the source plane near the tangential caustic would improve the ability of
the model to match these secondary peaks. The new model has six new parameters:
the flux of the second source ($F_{\rm s2}$), the position of the second source
relative to the SMA emission centroid ($\Delta \alpha_{\rm s2}$ and $\Delta
\delta_{\rm s2}$, constrained to be within $0\farcs2$ of the tangential
caustic), half-light radius of the second source ($r_{\rm s2}$), the ellipticity
of the second source ($\epsilon_{\rm s2}$), and the position angle of the second
source ($\phi_{\rm s2}$).  Initial tests of the two-source model showed that
the fitting routine struggled to identify the best-fit model consistently
unless the S\'ersic index of the second source was fixed at $n_{\rm s2} = 0.5$
(i.e., a Gaussian profile), so that is what we have adopted here.

\begin{figure*}[!tbp] 
\epsscale{1.00} 
\includegraphics[width=\textwidth]{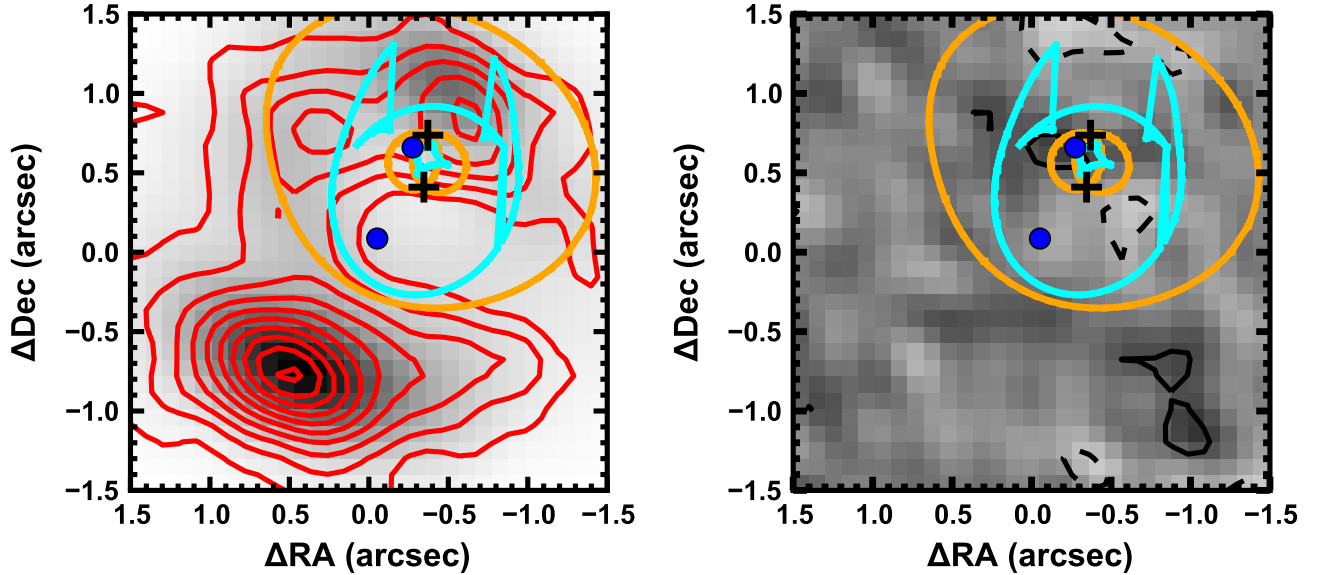}

\caption{ Same as Figure~\ref{fig:modeling}, but now with a two-source model
(filled blue circles indicate the peak flux positions of the two galaxies in
the source plane).  This model has a magnification factor of $\mu = 3.7\pm0.3$
and Einstein radii of $\theta_{\rm E1} = 0\farcs57\pm0\farcs02$ and
$\theta_{\rm E2} = 0\farcs40\pm0\farcs01$.  This suggests that the measurement
of these important parameters is relatively insensitive to whether a
single-source or two-source model is assumed.  \label{fig:merger}}

\end{figure*}

Figure~\ref{fig:merger} shows the results obtained with the two-source model
and Table~\ref{tab:lensingresults} contains the best-fit model parameters and
their 1$\sigma$ uncertainties. The best-fit model has $\chi^2 = 217769.2$ and
144185 degrees of freedom. The magnification factor is $\mu = 3.7 \pm 0.3$, and
the Einstein radii of the two lenses are $\theta_{\rm E1} = 0.57 \pm 0.01$ and
$\theta_{\rm E2} = 0.40 \pm 0.01$, values that are very similar to those
obtained with the single-source model. These are the parameters of greatest
interest, so it is reassuring that they are relatively insensitive to the exact
morphology of the background source.

The position of the primary background source is also relatively robust between
the two models we have tested, with ($\Delta \alpha_{\rm s} = 0\farcs05 \pm
0\farcs02$, $\Delta \delta_{\rm s} = 0\farcs09 \pm 0\farcs03$) relative to the
SMA emission centroid. The secondary source is less well-constrained, having a
position of $\Delta \alpha_{\rm s2} = 0\farcs2 \pm 0\farcs1$, $\Delta
\delta_{\rm s2} = 0\farcs53 \pm 0\farcs09$.  This position places the secondary
source near a caustic, implying that it has experienced a high degree of
magnification and has a much lower intrinsic luminosity than the primary
source.  The primary source maintains a broad profile intermediate between a
disk and an elliptical, while the second source is compact ($r_{\rm s2} = 2.2
\pm 0.8$~kpc). In both the single-source and two-source models, the best-fit
residual maps show emission to the southwest in the SMA data that the model
fails to reproduce. This could be an indication of the presence of a third
source in the source plane, but an examination at that level is beyond the
scope of this paper. Overall, the agreement between the single-source and
two-source model results is encouraging. For the remainder of the paper, we use
the results from our single-source model, since it provides nearly as good a
fit to the data as the two-source model ($\Delta \chi^2 = 29.9$) but requires
six fewer free parameters. Our major conclusions are insensitive to whether the
single-source or two-source model is used.

It is interesting that the magnification factor we have measured is somewhat
lower than might be expected based on the velocity dispersion and observed
luminosity of CO emission lines in G15v2.779.  Studies of CO($J=1-0$) emission
lines in both lensed and unlensed SMGs have found a correlation between the
intrinsic (i.e., unlensed) CO($J=1-0$) line luminosity and the FWHM of the
emission line (Harris et al., submitted; Bothwell et al., in prep.).  While
CO($J=1-0$) data are not yet available for G15v2.779, higher $J$ lines have
indicated FWHM$= 700~$km$\;$s$^{-1}$ and $L_{{\rm CO}(J=5-4)}^\prime = 3 \times
10^{11} ~K \;$km$\,$s$^{-1}\;$pc$^2$.  According to the correlation identified
in the previous studies, this FWHM value would imply an (unlensed) line
luminosity of $L_{{\rm CO}(J=1-0)}^\prime = 7 \times 10^{10} ~K
\;$km$\,$s$^{-1}\;$pc$^2$.  However, assuming a typical sub-thermal CO($J=5-4$)
to CO($J=1-0$) conversion factor (e.g., Harris et al., in submitted.), our
measurement of $\mu$ would indicate $L_{{\rm CO}(J=1-0)}^\prime = 1.7 \times
10^{11} ~K \;$km$\,$s$^{-1}\;$pc$^2$.  Possible explanations for this include
either an unusual CO($J=5-4$)/CO($J=1-0$) ratio, a large intrinsic scatter in
the luminosity line-width relations in these systems, or a difference in the
size-scale of the CO($J=1-0$) and the far-IR emitting regions, leading to
$\mu_{\rm CO}$ not being equal to $\mu_{\rm far-IR}$.

\section{The Nature of G15v2.779}\label{sec:nature}

In this section, we discuss the implications of our results for the nature of
G15v2.779.  We begin with a focus on the background source and end with the
foreground lens properties.

\subsection{The Background Source: An optically-obscured SMG at
$z=4.243$}\label{sec:bg}

The magnification factor of the background source is a parameter that is
critical to the derivation of any intrinsic property of the source.  The
properties that we consider here are the total IR luminosity ($L_{\rm IR}$),
the SFR, the projected IR luminosity surface density ($\Sigma_{\rm IR}$), the
dust mass ($M_{\rm dust}$), and the molecular hydrogen gas mass ($M_{\rm
H_2}$).  We compute $L_{\rm IR}$ as the integral under the full SED as reported
in \citet{Cox:2011fk}, divided by $\mu$.  Differential magnification effects
are an important consideration when computing $L_{\rm IR}$.  Initial studies
indicate that when the magnification factor at any given wavelength is modest
($\mu < 5$), then differential magnification effects are expected to have a
minimal influence on the inferred IR luminosity \citep{Serjeant:2012lr}.  The
contribution from the lensing galaxies to $L_{\rm IR}$ is likely to be minimal,
given that they are both early-type galaxies at $z=0.59$.  We find an intrinsic
IR luminosity of $L_{\rm IR} = (2.1\pm0.2) \times 10^{13}~L_\sun$, a value that
likely makes G15v2.779 one of the intrinsically brightest known SMGs (with an
intrinsic, un-lensed sub-mm flux density of $F_{\rm 880 \mu m} ({\rm
intrinsic}) = 21\pm2~$mJy).  

Using the standard conversion from $L_{\rm IR}$ to SFR
\citep{1998ARA&A..36..189K}, this corresponds to a SFR of
$3000\pm300~M_\sun~$yr$^{-1}$.  This value assumes that all of the IR
luminosity originates from star-formation rather than AGN.  Hence this value
should be regarded as an upper limit on the true SFR, although it should be
noted that measurements of the radio luminosity suggest a far-IR to radio flux
ratio that is consistent with starburst dominated galaxies
\citep{Cox:2011fk}.

The S\'ersic index of the background source ($n_{\rm s} = 2.9 \pm 0.3$) is
intermediate between an exponential disk profile ($n_{\rm s} = 1$) and a
De~Vaucouleurs ($n_{\rm s} = 4$) profile.  The source is significantly
extended, having a half-light radius of $4.4\pm 0.5~$kpc.  This implies a
de-projected IR luminosity surface density (computed as $\Sigma_{\rm IR} =
0.5L_{\rm IR} / A_{\rm half}$, where $A_{\rm half} = \pi r_{\rm s}^2$) of
$\Sigma_{\rm IR} = (3.4\pm0.9) \times 10^{11} L_\sun \;$kpc$^{-2}$.
This range of values for $\Sigma_{\rm IR}$ places G15v2.779 within the ``normal
star-forming'' class of galaxies at $z > 2$ and is significantly below that of
local ULIRGs \citep[which have $\Sigma_{\rm IR}$ values in the range $10^{12} -
10^{14} L_\sun \;$kpc$^{-2}$;][]{Rujopakarn:2011dq}.  This may be a clue that
unlike local ULIRGs, the intense, dust-enshrouded burst of star-formation that
is occurring in G15v2.779 may not be driven by a major merger.

To compute the apparent (i.e., un-corrected for lensing) dust mass,
\citet{Cox:2011fk} perform a single-temperature, optically thin, modified
black-body fit to the far-IR and sub-mm SED of G15v2.779
\citep[following][]{2006ApJ...642..694B}.  These authors find a dust mass of
$M_{\rm dust} = 8.9 \times 10^9 \mu^{-1} \; M_\sun$, using the best-fit dust
temperature of $T_{\rm dust} = 38~$K and a mass absorption coefficient of
$\kappa = 0.4~$cm$^2\;$g$^{-1}$.  If instead an optically thick modified
black-body is used to fit the data, the inferred dust mass decreases by nearly
a factor of 2 \citep{2007A&A...467..955W}.  Our measurement of $\mu$ implies
dust masses $M_{\rm dust} \sim 2 \times 10^9~M_\sun$ \citep[using the values
reported in][]{Cox:2011fk}.  The uncertainty in the dust mass is dominated by
systematic uncertainties related to the unknown optical depth at far-IR
wavelengths and the unconstrained mass opacity coefficient and is at least a
factor of a few.  The measured dust mass is similar (though at the top end) to
those estimated for other high-$z$ SMGs \citep{2010A&A...518L.154S} and also
for the most massive $z=0.5$ galaxies in H-ATLAS \citep{Dunne:2011lr}.

To quantify the interstellar medium properties in this object,
\citet{Cox:2011fk} use a spherical, single-component, large velocity gradient
(LVG) model \citep{2007A&A...467..955W} to fit simultaneously the CO($J=7-6$),
CO($J=5-4$), and CO($J=4-3$) emission lines.   Assuming a conversion factor of
$\alpha_{\rm CO} = 0.8 ~M_\sun \; ($K$\;$km$\;$s$^{-1}\;$pc$^2)^{-1}$ to go
from $L^\prime_{\rm CO(1-0)}$ to $M_{\rm H_2}$, \citet{Cox:2011fk} find gas
masses of $M_{\rm H_2} = 3.5 \times 10^{11}\; \mu^{-1} ~ M_\sun$.  Using our
measurement of $\mu$, we find $M_{\rm H_2} = (8\pm4) \times 10^{10}~M_\sun$.
Note that there is some evidence that $\alpha_{\rm CO}$ may be higher for less
dense systems at high-$z$ relative to local ULIRGS \citep{2011ApJ...726L..22F,
2011MNRAS.412.1913I}.  If this is indeed the case, then the gas mass would be
even larger.  Given the large dust mass (even with the optically thick fit),
this is not unreasonable.

This gas mass is a factor of $\approx 2.5$ larger than the typical gas mass
found in un-lensed SMGs \citep{2005MNRAS.359.1165G}, 2.5-9 times larger than
two lensed sources discovered in the H-ATLAS SDP \citep{2011ApJ...726L..22F},
and a factor of $\approx 2.5$ larger than a lensed source from HerMES
\citep{2011ApJ...733...29S}.  G15v2.779 appears to be a very massive, highly
star-forming galaxy at $z = 4.243$.

Table~\ref{tab:smgproperties} summarizes the properties of G15v2.779 and
compares them with unlensed SMGs at $z \sim 4$.  G15v2.779 bears a close
resemblence to GN20, having similar $L_{\rm IR}$, $M_{\rm gas}$, $\sigma_{\rm
gas}$, $r_{\rm s}$, $M_{\rm dyn}$, and $\Sigma_{\rm IR}$ values.  However,
there is an important difference: GN20 is much more luminous in the rest-frame
optical than G15v2.779.  In terms of visual extinction ($A_V$), G15v2.779 is
most similar to GN10, which has $A_V \sim 5-7.5$.  However, GN10 is less
luminous and has only an upper limit on its source size.  The fact that
G15v2.779 is clearly extended on scales $> 2$~kpc and yet maintains a covering
fraction near unity indicates an impressive quantity of dust, consistent with
the dust mass measurements described above.  

One feature all of these $z \sim 4$ SMGs share in common is $\Sigma_{\rm IR}$
values that are 1-2 orders of magnitude lower than those of ULIRGs in the local
Universe.  This may be an indication that the physical mechanisms driving the
prodigious luminosities in these systems may be different from what occurs at
$z \sim 0$.  Theories attempting to explain the behavior of these systems must
also account for the short gas depletion timescales \citep[possibly via strong
inflow of gas from the inter-galactic medium; e.g.][]{Schaye:2010fj}, as the
estimated SFRs will consume all of the available gas within $\sim 10-30$~Myr in
all of these galaxies.

\begin{centering}
\begin{deluxetable*}{lcccccccccc}
\tabletypesize{\scriptsize} 
\tablecolumns{11}
\tablewidth{500pt}
\tablecaption{Properties of $z \gtrsim 4$ SMGs}
\tablehead{
\colhead{} & 
\colhead{$L_{\rm IR}$} & 
\colhead{SFR} & 
\colhead{} & 
\colhead{$M_{\rm gas}$} & 
\colhead{$M_{\rm stars}$} & 
\colhead{} &
\colhead{$\sigma_{\rm gas}$} &
\colhead{$r_{\rm s}$} &
\colhead{$M_{\rm dyn} \,{\rm sin}^2i$} &
\colhead{$\Sigma_{\rm IR}$} \\
\colhead{} & 
\colhead{($10^{13}\;L_\sun$)} & 
\colhead{($M_\sun \;$yr$^{-1}$)} & 
\colhead{$f_{\rm AGN}$} & 
\colhead{($10^{11}\;M_\sun$)} & 
\colhead{($10^{11}\;M_\sun$)} & 
\colhead{$A_V$} &
\colhead{(km$\;$s$^{-1}$)} &
\colhead{(kpc)} &
\colhead{($10^{11}\;M_\sun$)} &
\colhead{($10^{12}\;L_\sun\;$kpc$^{-2}$)}
}
\startdata
G15v2.779   & $2.1\pm0.2$ & $3000\pm300$ & low\tnm{f}\tnm{g} & $0.8\pm0.4$ & --- & $>4$ & $800\pm100$ & $4.4\pm0.5$ & $3\pm1$ & $0.34\pm0.09$  \\
GN10\tnm{a} & $1.2\pm0.6$ & $1700\pm800$ & low\tnm{g} & $0.27\pm0.05$ & $1.0\pm0.5$ & $5-7.5$ & $770\pm200$ & $<4$ & $< 2.5$ & $> 0.1$ \\
GN20\tnm{b} & $2.9\pm1.6$ & $4300\pm2000$ & low\tnm{g} & $0.50\pm0.25$ & $2.3\pm1.5$ & $\lesssim 2$ & $710\pm120$ & $2\pm1$ & $2.3\pm1.5$ & $1.1\pm0.5$ \\
GN20.2a\tnm{b} & $1.6\pm1.0$ & $2300\pm1100$ & low\tnm{g} & $0.30\pm0.15$ & $0.5\pm0.3$ & $\lesssim 2$ & $1100\pm400$ & $<4$ & $<5$ & $> 0.1$ \\
J1000+0234\tnm{c} & $1.2\pm0.7$ & $1700\pm900$ & low\tnm{g} & $0.26\pm0.13$ & $0.30\pm0.15$ & $1.4\pm0.5$ & $600\pm200$ & $3.5\pm2.0$ & $1.3\pm0.7$ & $0.2\pm0.1$ \\
J033229.4\tnm{d} & $0.6\pm0.3$ & $900\pm450$ & low\tnm{h} & $0.16\pm0.03$ & $<0.5$ & $1.5\pm0.5$ & $160\pm65$ & $2\pm1$ & $0.12\pm0.06$ & $0.2\pm0.1$ \\
AzTEC-3\tnm{e} & $1.7\pm0.8$ & $1800\pm900$ & low\tnm{g} & $0.53\pm0.25$ & $0.10\pm0.05$ & $\lesssim 2$ & $487\pm58$ & $<5$ & $>1.4$ & $>0.1$ \\
\tnt{a}{\citet{Daddi:2009kx}}
\tnt{b}{\citet{Daddi:2009qy, Carilli:2011qy}}
\tnt{c}{\citet{Capak:2008lr, Schinnerer:2008uq}}
\tnt{d}{\citet{Coppin:2009lr,Coppin:2010zr}}
\tnt{e}{\citet{Riechers:2010fk}}
\tnt{f}{880$\mu$m size $> 2$~kpc}
\tnt{g}{Radio to IR luminosity ratio consistent with star-formation dominated galaxies}
\tnt{h}{Weak NV, 24$\mu$m, and near-UV emission inconsistent with AGN}
\enddata
\label{tab:smgproperties}
\end{deluxetable*}
\end{centering}

\subsection{The Foreground Lenses: A Dry Merger at $z=0.595$}\label{sec:fg}

We use the standard equations from \citet{Schneider:1992fk} to compute the mass
of the lens galaxies $M_{\rm lens}$, finding $M_{\rm lens1} = (7.4\pm0.5)
\times 10^{10} ~ M_\sun$ (recall that we have assumed $M_{\rm lens1} = 2 \times
M_{\rm lens2}$).  Independent mass estimates can be obtained using the
correlation between $V$-band luminosity and $M_{\rm lens}$ given in
\citet{Negrello:2010fk}.  At $z_{\rm lens} = 0.59$, the observed $z$-band
corresponds almost exactly to the rest-frame $V$-band.  The ground-based
$z$-band magnitude of the two lenses together is $z = 20.35 \pm 0.40$
\citep{Cox:2011fk}, corresponding to a rest-frame $V$-band luminosity of $\nu
L_\nu (V) = 3.3^{+1.5}_{-1.1} \times 10^{10}~L_\sun$.  Assuming the rest-frame
$V$-band luminosity ratio between the two lens galaxies is the same as that in
the observed $K_{\rm S}$-band, the $L_V - M_{\rm lens}$ correlation observed
from the SLACS lenses \citep{Bolton:2008wd, Negrello:2010fk} implies lens
masses of $M_{\rm lens1} = (4\pm2) \times 10^{10}~M_\sun$ and $M_{\rm lens2} =
(2\pm1) \times 10^{10}~M_\sun$, consistent with our lens model estimates of
$M_{\rm lens1}$ and $M_{\rm lens2}$.  An alternative estimate of the lens
masses is to use the $V$-band luminosities along with a mass-to-light ratio
derived from synthesized stellar populations.  The optical spectrum of
G15v2.779 shows a strong Balmer/4000 \AA\ break ($D_n\;(4000) = 2.0\pm0.1$),
typical of galaxies dominated by old stellar populations \citep[5-10~Gyr,
depending on metallicity;][]{Kauffmann:2003qf}.  A synthesized stellar
population with an age of 5~Gyr, a Chabrier~IMF, and no dust extinction has a
mass-to-light ratio of $M_{\rm star} / \nu L_\nu \,(V) \approx 2$.  Thus, the
inferred stellar masses of the two lensing galaxies are $M_{\rm star,1} \approx
(4\pm2) \times 10^{10} \, M_\sun$ and $M_{\rm star,1} \approx (2\pm1) \times
10^{10} \, M_\sun$.  
There is evidence at the 1.5$\sigma$ level that the lens and mass estimates are
greater than the stellar mass estimates.  This may be reasonable considering
the small stellar sizes (half-light radii of $\approx 0.9$~kpc) relative to the
Einstein radius of each system ($\sim 3$~kpc).  

\begin{figure}[!tbp] 
\begin{centering}
\includegraphics[width=0.45\textwidth]{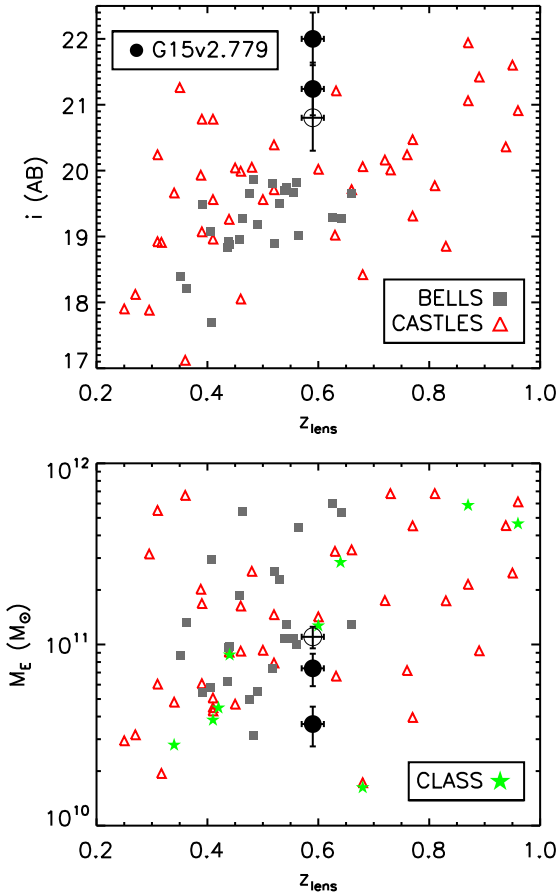}

\caption{ Comparison of lens properties from BELLS \citep[gray
squares;][]{Brownstein:2012cr}, CASTLeS \citep[red triangles;][]{Munoz:1998mz},
CLASS \citep[green stars;][]{Myers:2003lr}, and G15v2.779 (black filled circles
represent individual lens galaxies; open circle represents the sum of both lens
galaxies).  {\it Top}: Lens $i$-band magnitudes as a function of redshift.
Note that G15v2.779 is significantly fainter in $i$-band than any of the
galaxies in BELLS.  {\it Bottom}: Mass within the Einstein radius as a function
of redshift.  G15v2.779 is less massive than all BELLS galaxies at $z_{\rm
lens} > 0.5$.  This demonstrates that as a source-selected lens survey (similar
to CLASS and part of CASTLeS), H-ATLAS will be less biased towards massive,
bright lensing galaxies than lens-selected surveys like SLACS or BELLS.
H-ATLAS and other {\it Herschel} wide-field surveys will help identify lens
systems at intermediate redshifts where the galaxy in the foreground is
particularly faint, either due to having a low mass or lieing at high redshift
(or both, as appears to be the case for G15v2.779).  \label{fig:castlesboss}}
\end{centering}

\end{figure}

Figure~\ref{fig:castlesboss} shows that G15v2.779 has two of the faintest (in
$i$-band) and lowest mass (at this redshift) lensing galaxies found in current
surveys of gravitationally lensed systems \citep[e.g., CASTLeS, CLASS, SLACS,
and BELLS;][]{Munoz:1998mz, Myers:2003lr, Bolton:2008wd, Brownstein:2012cr}.
This is an indication that wide-field surveys with {\it Herschel} will be
useful for identifying lensing systems where the lensing galaxy is faint in the
optical, either due to being low-mass or being very distant.  This is generally
true for source-selected lensing surveys (e.g., CLASS, H-ATLAS, and a portion
of CASTLeS), whereas lens-selected lensing surveys (e.g., SLACS and BELLS) tend
to be biased towards brighter and more massive foreground galaxies.

Both G15v2.779 lens galaxies have early-type morphologies with small half-light
radii ($\approx 0.9$~kpc).  Such systems appear to be commonplace at $z \sim
2$, but become increasingly rare at lower redshifts
\citep[e.g.,][]{Damjanov:2009nx}.  However, given the small separation between
the two lensing galaxies ($\approx 2$~kpc), it is likely that they are about to
merge together \citep[dissipationless mergers often do not have obvious signs
of interaction even at these separations; e.g., see][]{Bell:2006kx}.  Simple
virial arguments suggest that this process could lead to a doubling of the
radius while the mass only increases by 50\% \citep{Naab:2009oq}.  Such a
result would make the size of the merged system consistent with similarly
massive galaxies at $z \approx 0.5$ \citep[e.g.,][]{Oser:2012ul} as well as
more typical of lensing galaxies found in CASTLES, CLASS, SLACS, and BELLS.
Overall, these observations are consistent with the dissipationless (``dry'')
merging stage of the two-phase galaxy evolution scenario outlined in
\citet{Oser:2010eu}.  In this picture, the progenitors of massive galaxies
undergo intense in-situ star-formation from $z \sim 6$ to $z \sim 2$ that leads
to compact, elliptical galaxies with little or no reservoirs of gas for future
star-formation.  From $z \sim 2$ to the present-day, massive galaxies undergo
dry merging and increase their sizes such that they evolve onto the mass-size
relation observed in local early-type galaxies.

\section{Conclusions} \label{sec:conclusions}

We use high-spatial resolution imaging obtained with the SMA at 880$\mu$m and
Keck AO at $K_{\rm S}$-band to perform a detailed gravitational lens modeling
of G15v2.779, an SMG at $z = 4.243$ identified by {\it Herschel} in the H-ATLAS
survey.  We present a Gemini GMOS-S optical spectrum of G15v2.779 that
suggests that the two foreground galaxies are at $z_{\rm lens} =
0.595\pm0.005$.  This analysis provides important measurements of the nature of
both the background SMG and the foreground lenses.  We summarize our findings
below.

We employ a visibility-plane lens modeling analysis and find a magnification
factor of $\mu = 4.1\pm0.2$ for the background source.  This measurement is
significantly lower than what had been previously assumed for this source and
indicates that not all of the brightest lens candidates identified by {\it
Herschel} have high magnification factors.  This value of $\mu$ implies an
intrinsic infrared luminosity of $L_{\rm IR} = 2.1\pm0.2 \times
10^{13}~L_\sun$.  

The best-fit model for the background source favors radial profiles that are
intermediate between exponential disks and De Vaucouleurs.  The half-light
radius of the background source is $r_{\rm s} = 4.4\pm0.5 \;$kpc.  This size
measurement implies a de-projected IR luminosity surface density of
$\Sigma_{\rm IR} = (3.4\pm0.9) \times 10^{11} L_\sun \;$kpc$^{-2}$.  This
number is typical of $z > 2$ ULIRGs and HyLIRGs but 10-100 times lower than
ULIRGs in the local Universe.  This may be an indication that the formation
mechanism for this source could be different from $z \sim 0$ ULIRGs, which are
thought to arise from major mergers of gas-rich disk galaxies.  Higher-spatial
resolution data with improved sensitivity are needed to favor one of these
models over the other.

Our measurement of $\mu$, in conjunction with previous observations of CO
emission lines and the far-IR SED, indicates a gas mass of $M_{\rm H_2} \approx
(8\pm4) \times 10^{10}~M_\sun$ and a dust mass of $M_{\rm dust} \sim 2 \times
10^9~M_\sun$.  These values are factors of 2.5-9 times larger than other lensed
galaxies studied to date but are comparable to other $z \sim 4$ SMGs.  They
indicate that G15v2.779 hosts a massive reservoir of molecular gas that is
fueling a prodigious, but likely short-lived ($\sim 10-30$~Myr) period of
star-formation.

The foreground lenses have Einstein radii of $\theta_{\rm E1} = 0\farcs57 \pm
0\farcs01$ and $\theta_{\rm E2} = 0\farcs40 \pm 0\farcs01$.  These imply lens
masses of $M_{\rm lens1} = (7.4\pm0.5) \times 10^{10} ~ M_\sun$ and $M_{\rm
lens2} = (3.7\pm0.3) \times 10^{10} ~ M_\sun$.  The lensing galaxies have sizes
of $\approx 0.9\;$kpc and lie at a redshift of $z=0.595 \pm 0.005$.  The
gravitational potential from both galaxies may include a significant
contribution from dark matter.  They are separated by 2~kpc and will likely
merge into a single early-type galaxy with a larger size that will make the
resultant system consistent with other early-types at $z \sim 0.6$.  

Together, the SMA, Keck, and Gemini data have established that G15v2.779 is a
SMG at $z = 4.243$ modestly lensed by a pair of early-type galaxies at
$z=0.595$.  Our results highlight the bounty of information that can be obtained
via a multi-wavelength approach to studying strongly lensed SMGs at high
redshift.  More sensitive and higher-spatial resolution imaging of the lensed
emission is needed to improve the constraints on the parameters of the
gravitational lensing model and test competing models for the powering
mechanism in this source (e.g., major merger vs. secular processes).  This will
become feasible in the near future when baseline lengths of $\approx 1$~km
become available with ALMA.  {\it HST} spectroscopy is needed to confirm that
the lensing galaxies both lie at $z=0.59$, to measure their velocity
dispersions, and to improve their stellar mass estimates.  

\begin{acknowledgments}

The results described in this paper are based on observations obtained with
{\it Herschel}, an ESA space observatory with science instruments provided by
European-led Principal Investigator consortia and with important participation
from NASA.  The {\it Herschel}-ATLAS is a project with {\it Herschel}, which is
an ESA space observatory with science instruments provided by European-led
Principal Investigator consortia and with important participation from NASA.
The H-ATLAS website is http://www.h-atlas.org/.  US participants in H-ATLAS
acknowledge support from NASA through a contract from JPL.  RSB acknowledges
support from the SMA Fellowship program. HF, AC, JLW and SK acknowledge support
from NSF CAREER AST-0645427.  We thank the referee for a thorough review
of the manuscript which resulted in a stronger paper overall.

The ground-based follow-up observations were obtained at the SMA, at the
W.~M.~Keck Observatory, and at the Gemini South Observatory.  The SMA is a
joint project between the Smithsonian Astrophysical Observatory and the
Academia Sinica Institute of Astronomy and Astrophysics and is funded by the
Smithsonian Institution and the Academia Sinica. The authors wish to recognize
and acknowledge the very significant cultural role and reverence that the
summit of Mauna Kea has always had within the indigenous Hawaiian community.
We are most fortunate to have the opportunity to conduct observations from this
mountain.  

Based on observations obtained at the Gemini Observatory, which is operated by
the Association of Universities for Research in Astronomy, Inc., under a
cooperative agreement with the NSF on behalf of the Gemini partnership: the
National Science Foundation (United States), the Science and Technology
Facilities Council (United Kingdom), the National Research Council (Canada),
CONICYT (Chile), the Australian Research Council (Australia), Minist\'erio da
Ci\^encia, Tecnologia e Inova\c{c}\~ao (Brazil) and Ministerio de Ciencia,
Tecnolog\'ia e Innovaci\'on Productiva (Argentina).  Facilities: SMA, Keck,
Gemini-South.

\end{acknowledgments}



\end{document}